\documentstyle[aps,twocolumn,epsf]{revtex}
\begin{document}
\title{On the classical character of control fields in quantum information processing}
\author{S.J. van Enk$^1$ and H.J. Kimble$^2$\\
$^1$Bell Labs, Lucent Technologies,
Room 2C-401\\
600-700 Mountain Ave,
Murray Hill NJ 07974\\
$^2$Norman Bridge Laboratory of Physics 12-33\\
California Institute of Technology,
Pasadena CA 91125}
\maketitle
\abstract{Control fields in quantum information processing are virtually always, almost by definition, assumed to be classical. In reality, however, when such a field is used to manipulate the quantum state of qubits, the qubits never remain completely unentangled with the field. For quantum information processing this is an undesirable property, as it precludes perfect quantum computing and quantum communication. Here we consider the interaction of atomic qubits with laser fields and quantify atom-field entanglement in various cases of interest.
We find that the entanglement decreases with the average number of photons $\bar{n}$ in a laser beam as $E\propto\log_2 \bar{n}/\bar{n}$ for $\bar{n}\rightarrow\infty$.} 
\section{Introduction}
In many protocols for the implementation of quantum logic, auxiliary
control fields are employed to address small quantum systems individually.
These control fields are almost universally assumed to be classical. For
instance, an important group of physical implementations for quantum computing and quantum communication involves laser fields and single matter particles, such as electrons, atoms or atomic ions (for a recent overview see \cite{fort}). The latter can be used to store quantum information in different spin states or ground and metastable excited states, and the former allow one to apply the desired one-bit or multi-bit operations. Even the relatively simple one-bit operations, however, are never perfect in practice. Ref.~\cite{ion} provides an extensive discussion 
of sources of imperfections of one-bit operations in the ion-trap quantum computer \cite{cz,steane,hughes}. Such imperfections, often referred to as decoherence in the context of quantum-information processing, can always be attributed to some unwanted but often unavoidable interaction of the quantum system with its environment. For example, spontaneous emission is due to the interaction of an atom or ion with the empty modes of the radiation field. Other decoherence effects considered in Ref.~\cite{ion} are due to collisions of the ions with a background gas, interactions with the walls of the trap, and interactions with fluctuating electric and magnetic fields. 

One decoherence effect not considered in Ref.~\cite{ion} and elsewhere, is the fact that 
the laser field will inevitably become entangled with the atom. 
This is simply because the transition from ground state to excited state (or {\em vice versa}) is accompanied by the absorption (or emission) of a laser photon. If the absence of a laser photon or the presence of an extra photon is in principle detectable, then there is information about the state of the qubit present in the environment (in this case, the laser field), which leads to decoherence.
In general one expects such decoherence effects to diminish with increasing photon numbers because of the correspondence principle; for a classical field there would be no entanglement. 
Of course, if one were to use highly nonclassical states of the radiation field, such as photon number states, then this expectation would not be fulfilled, but for a laser beam well described by a mixture of number states with a Poissonian probability distribution, the entanglement indeed decreases with the average number of photons, as we will show here.

One important parameter that determines the amount of atom-field entanglement in a given experiment is the focal area of the light beam.
For instance, in an ion-trap quantum computer
containing several ions each
ion can in principle be addressed by focusing a laser beam onto the
appropriate position. The focusing requirements are then obviously
determined by the distances between neighboring ions. The same would apply
to the situation where several atoms are kept inside optical cavities
\cite{ye}, for the purpose of quantum computation
\cite{pell95} or communication
\cite{enk2}.
For a small array of qubits with
large spacings, the assumption of a classical laser field may indeed be justified. However, as the
density of qubits increases, the external control field must be focused
ever more tightly to avoid parasitic excitation of neighboring qubits. The
question then arises whether the assumption of a classical field
is justified for an atom localized on a wavelength scale with
illumination of large numerical aperture. With such localization and
illumination, the transmitted field might have imprinted upon it
measurable signatures of its interaction with the atom. Such entanglement
between atom and field would cause quantum information encoded in the atom
to decohere.
Of course there are avenues to
mitigate this difficulty, as for example by focusing in a cylindrical
geometry to increase the beam area while still keeping a small dimension
along a linear array of atoms. Perhaps surprisingly, the general solution
to this problem, e.g., for forward scattering and fluorescent
fields is not known, even for the simple case of light focused onto a
two-state atom. Relevant work includes the application of a standard input-output formalism
to a quasi 1-dimensional version of this problem \cite{car}, and the construction of exact 3-dimensional vector solutions of the Maxwell equations, representing
beams of light focused by a strong spherical lens\cite{focus},
but these calculations do not directly address the question of entanglement.
We attempt to fix that problem here.

We wish to assess the importance of decoherence (and its dependence on focusing parameters) due to atom-field entanglement under typical experimental conditions. We, therefore, do not consider any of the other decoherence mechanisms mentioned above, although we do briefly discuss spontaneous emission. We discuss atom-field entanglement in the case that one or two coherent fields (from an ideal laser) interact with a single atom.
We first consider in Sec.~\ref{quadru} the simple case of a {\em single} laser field connecting two states, one of which is a ground state, the other a metastable state so that spontaneous emission occurs only with a small probability.
The second case we discuss is that
of a two-photon Raman transition from one ground state to another ground state (Sec.~\ref{raman}). By detuning sufficiently far from the intermediate excited state one can again suppress spontaneous emission.
\section{Preliminaries}
\subsection{Atom-field entanglement}
We consider an atom and denote by $|0\rangle$ and $|1\rangle$ its two energy eigenstates that encode the quantum bit.   
The atom is assumed to be initially
in some pure state
\begin{equation}
|\psi(\theta,\varphi)\rangle=\cos(\theta/2)|0\rangle+\sin(\theta/2)\exp(i\varphi)|1\rangle.
\end{equation}
This is sufficiently general for our purposes.
 Then we would like to apply a one-bit operation such as a NOT operation or a $\sqrt{{\rm NOT}}$ operation to the atom by using a laser field. This will in general entangle the atom with that laser field and, moreover, with other systems that the laser is already entangled with, and, finally, also with other initially empty modes of the radiation field when there is spontaneous emission. In the calculations below we assume, without loss of generality, the quantum
state of the whole system, laser field plus atom plus environment, to be pure. 
We can then define the entanglement between the atom and the laser field plus environment
as a function of time $t$ by \cite{ent}
\begin{equation}\label{entg}
E_{\theta,\varphi}(t)=-{\rm Tr} \rho_{\theta,\varphi}(t) \log_2\rho_{\theta,\varphi}(t)
\end{equation}
with $\rho_{\theta,\varphi}(t)$ the reduced density matrix of the atom, obtained by tracing out the field and environment. The subscripts denote the dependence on the initial atomic state. We will be interested in the average entanglement $\langle E\rangle$, averaged over all initial states of the qubit,
\begin{equation}\label{enta}
\langle E\rangle(t)=\frac{1}{4\pi}\int_0^{2\pi} {\rm d}\varphi \int_0^{\pi} {\rm d}\theta \sin\theta E_{\theta,\varphi}(t).
\end{equation}
For ease of description we will refer to this quantity as the entanglement between the atom and the laser field.
\subsection{Laser fields and coherent states}
Even if the field inside a laser cavity can be approximated by a single-mode coherent state, the light emanating from the laser will contain a continuum of frequencies. Thus the quantum state of a laser pulse is more correctly described by a continuous-mode coherent state \cite{mixed}.
In order not to make the problem more complicated than it already is, we
will assume that the laser beam is well approximated by a one-dimensional light beam \cite{loudon} propagating in the positive $z$ direction with a well-defined polarization vector $\vec{\varepsilon}$. This is correct as long as the light beam is not focused too strongly (to areas of size $A\approx \lambda^2$) \cite{focus}.
We define continuous-mode creation and annihilation operators
$a^{\dagger}(\omega)$ and $a(\omega)$ for each frequency $\omega$, and write for the electric field
operator as a function of position $z$ 
\begin{equation}\label{1D}
\vec{{\cal E}}(z)=\int d\omega \sqrt{\frac{\hbar\omega}{4\pi\epsilon_0cA}}
\big[a(\omega)\vec{\varepsilon}\exp(i\omega z/c)+H.c.\big],
\end{equation} 
where $A$ is the area of the beam and $H.c.$ stands for Hermitian conjugate.
A continuous-mode coherent state can be defined as
\begin{equation}\label{coco}
|\alpha(\omega)\rangle=\exp\left(\int{\rm d}\omega [
\alpha(\omega) a^{\dagger}(\omega) -\alpha^*(\omega)a(\omega)]\right)|{\rm vac}\rangle,
\end{equation}
with $|{\rm vac}\rangle$ the vacuum state and $\alpha(\omega)$ the continuous-mode coherent state amplitudes.
The expectation value of the electric field operator in a freely evolving coherent state ---for which $\alpha(\omega)$ evolves in time as $\alpha_t(\omega)=\exp(-i\omega t)\alpha(\omega)$--- is equal to the corresponding time-dependent classical field, i.e.,
\begin{eqnarray}\label{clas}
\langle\alpha_t(\omega)|\vec{{\cal E}}(z)|\alpha_t(\omega)\rangle&=&
2 {\rm Re} \left[\int d\omega \sqrt{\frac{\hbar\omega}{4\pi\epsilon_0cA}}
\alpha(\omega)\vec{\varepsilon}e^{i\omega (z/c-t)}\right]\nonumber\\
&\equiv&\vec{{\cal E}}_{{\rm clas}}(z,t)
\end{eqnarray} 
\subsubsection{Discrete coherent states}\label{discrete}
An alternative description of the continuous-mode coherent state $|\alpha(\omega)\rangle$ makes use of {\em discrete} creation and annihilation opertors and corresponding discrete coherent states. Following \cite{loudon} we let ${\phi_i(t)}$ be a complete set of functions such that
\begin{eqnarray}\label{complete}
\int{\rm d}t\phi_i(t)\phi^*_j(t)&=&\delta_{ij},\nonumber\\
\sum_i \phi^*_i(t)\phi_i(t')&=&\delta(t-t').
\end{eqnarray}
In terms of this set one can define bosonic creation and annihilation operators
$c_i$ by
\[
c_i=\int {\rm d}t \phi^*_i(t)a(t),
\]
where $a(t)$ is the Fourier transform of the operator $a(\omega)$,
\[
a(t)=\frac{1}{\sqrt{2\pi}}\int{\rm d}\omega a(\omega)\exp(-i\omega t).
\]
The continuous-mode coherent state (\ref{coco}) can now be written as a tensor product of coherent states $\otimes_i|\gamma_i\rangle$, where $|\gamma_i\rangle$ is the eigenstate of the annihilation operator $c_i$ with eigenvalue
\[
\gamma_i=\int {\rm d}t \phi^*_i(t)\alpha(t),
\]
with $\alpha(t)$ the Fourier transform of $\alpha(\omega)$.
\subsection{Atom-light interaction}
We start with the simple case of a two-level atom irradiated by a single laser field.
The Hamiltonian describing a two-level atom positioned at $z=0$ in a continuous field, using the usual long-wavelength and rotating-wave approximations, and transforming to a frame rotating at the atomic frequency $\omega_0=(E_1-E_0)/\hbar$ is
\begin{equation}\label{H}
H=\int{\rm d}\omega \left[\hbar \Delta(\omega)
a^{\dagger}(\omega)a(\omega) 
+g(\omega)(\sigma^{\dagger}a(\omega) +
a^{\dagger}(\omega) \sigma^-)\right],
\end{equation}
with $\Delta(\omega)=\omega-\omega_0$ the detuning from atomic resonance, $\sigma^{\pm}$ the atomic raising and lowering operators, and
\begin{equation}
g(\omega)=d \sqrt{\frac{\hbar\omega}{4\pi\epsilon_0cA}}.  
\end{equation}
The coupling constant $d$ is assumed real. In the usual case of a dipole transition $d$ would be the atomic dipole moment. For a quadrupole transition to a metastable state the leading interaction term contains the atomic quadrupole moment $Q$ and the gradient of the electric field. The same Hamiltonian (\ref{H}) is valid where now $d=2\pi Q/\lambda$, with $\lambda$ the wavelength of the laser light.

The evolution operator $U(t)=\exp(-iHt/\hbar)$ is not easily evaluated explicitly in the general case. However, if the bandwidth $B$ (the spread of frequencies) of the field is sufficiently small, a condition specified below, we can approximate the Hamiltonian by that of an atom in a fictituous single-mode coherent state with one frequency which we denote by $\omega_L$. We tackle the problem of introducing the required approximations in two steps.

First consider the following simple Hamiltonian, 
\begin{equation}\label{JC}
\tilde{H}=\hbar \Delta a^{\dagger}a+\hbar g(a^{\dagger}\sigma^-+\sigma^{\dagger}a),
\end{equation} 
with $\Delta=\omega_L-\omega_0$ the detuning from atomic resonance and $a$ and $a^{\dagger}$ the annihilation and creation operators of the fictituous single-mode field. 
The Hamiltonian $\tilde{H}$ describes the well-known Jaynes-Cummings model.
In fact, using this model the entanglement of a two-level atom interacting with a single-mode quantized field was studied in the early 90s within a very different context, namely, the occurrence of so-called collapses and revivals on very long time scales \cite{phoenix}. Here we are rather interested in short time scales. In fact the Jaynes-Cummings model would not even be valid in our case for longer times.
For the Hamiltonian (\ref{JC}) an analytical solution of the evolution operator can be found easily \cite{scully}. In particular, expanding the time-dependent atom-field wave function as
\begin{eqnarray}
|\Psi(t)\rangle=\sum_n c_0^n(t) |n\rangle |0\rangle +c_1^n(t) |n\rangle |1\rangle 
\end{eqnarray}
we get
\begin{eqnarray}\label{sol}
c_1^n(t)=\left( \left[\cos (\Omega_n t/2)-\frac{i\Delta}{\Omega_n}\sin(\Omega_nt/2)\right] c_1^n(0)\right.\nonumber\\
\left.-\frac{2ig\sqrt{n+1}}{\Omega_n}\sin (\Omega_n t/2)c_0^{n+1}(0)
\right)\exp(i\Delta t/2),\nonumber\\
c_0^{n+1}(t)=\left( \left[\cos (\Omega_n t/2)+\frac{i\Delta}{\Omega_n}\sin(\Omega_nt/2)\right] c_0^{n+1}(0)\right.\nonumber\\
\left.-\frac{2ig\sqrt{n+1}}{\Omega_n}\sin (\Omega_n t/2)c_1^{n}(0)
\right)\exp(-i\Delta t/2),
\end{eqnarray}
with 
\begin{equation}
\Omega_n=\sqrt{\Delta^2+4g^2(n+1)}.
\end{equation}
From this solution we can read off the conditions under which the approximation of a laser field as a single-mode field is justified. Namely, a change in $\Delta$ by an amount $B$ should not change this solution appreciably.
This implies, first, that over the time period of interest, $T$, the change in phase $\Delta t$ is small, so that
\begin{equation}\label{cond2}
B T\ll 2\pi.
\end{equation}
The second condition is that $\Omega_n$ be affected only negligibly, so that
\begin{eqnarray}
B\ll g\sqrt{\bar{n}+1},
\end{eqnarray}
with $\bar{n}=|\alpha|^2$ the average number of photons in the single mode.

The Jaynes-Cummings Hamiltonian (\ref{JC}) describes an interaction with a field that does not change in time, whereas the original Hamiltonian (\ref{H}) can describe interactions with finite laser pulses. In order to introduce such a finite interaction into the Jaynes-Cummings Hamiltonian,  
we define a dimensionless function $\phi(t)$, a slowly varying envelope function describing the turning on and off of the atom-field interaction, by
\begin{equation}
\phi(t)=\frac{{\cal E}^{(+)}_{{\rm clas}}(z=0,t)}{{\rm max}_t {\cal E}^{(+)}_{{\rm clas}}(z=0,t)},
\end{equation}
where ${\cal E}^{(+)}_{{\rm clas}}$ is the positive-frequency component of the classical field 
(\ref{clas}) along the polarization vector.

In order to define in a consistent way the mode operators $a$ and $a^{\dagger}$, the amplitude $\alpha$, and the coupling constant $g$ we wish to make use of the formalism of Section \ref{discrete}. 
We can indeed construct a complete set of functions $\phi_i(t)$ satisfying (\ref{complete}) of which $\phi_1(t)\equiv \phi(t)/\sqrt{{\cal T}}$ is one member [the other members correspond to orthogonal modes that are initially empty], with
\[{\cal T}\equiv \int_{-\infty}^{\infty} {\rm d}t |\phi(t)|^2.\]
We then identify 
\begin{eqnarray}
a&\equiv& c_1=\int{\rm d}t\phi_1^*(t)a(t),\nonumber\\
\alpha&=&\int {\rm d}t \phi_1^*(t)\alpha(t).
\end{eqnarray}
With the function $\phi(t)$ describing the finite character of the atom-field interaction, the more accurate description of an atom interacting with a time-dependent pulse is by means of a time-dependent Hamiltonian
\begin{equation}\label{Ht}
\tilde{H}(t)=\hbar\Delta c_1^{\dagger}c_1+ \phi(t)\hbar g(c_1^{\dagger}\sigma^-+\sigma^{\dagger}c_1).
\end{equation} 
In this way we can fulfil the simple requirement that the expectation values in the state $|\alpha(\omega)\rangle$ of the interaction terms from the approximate Hamiltonian (\ref{Ht}) and  the actual Hamiltonian (\ref{H}) be the same. This in fact 
determines the effective coupling constant $g$,
\begin{equation}
\hbar g\alpha=d \max_t E^{(+)}_{{\rm clas}}(z=0,t).
\end{equation}
For narrow bandwidth $B$, one can write \cite[(3.10)]{loudon} 
\begin{equation}
\alpha(\omega)=\sqrt{2\pi F}\delta(\omega-\omega_L)\exp(i\theta),
\end{equation}
with $F=\langle a(t)a^{\dagger}(t)\rangle$  the flux and $\theta$ an arbitrary phase, which can be put equal to zero without loss of generality. 
This gives
\begin{equation}\label{ga}
\hbar g\alpha=\sqrt{\frac{\hbar \omega_L Fd^2}{2\epsilon_0 cA}}.
\end{equation}
For on-resonance excitation ($\Delta=0$) we can analytically solve the evolution equations by
defining a time variable 
\begin{equation}
\tilde{t}=\int_{-\infty}^t {\rm d}\tau \phi(\tau).
\end{equation}
The solution for atom and field takes the form (\ref{sol}) but with $t$ replaced by $\tilde{t}$. The total interaction time $T$ of the laser pulse is defined as
\begin{equation}
T=\int_{-\infty}^\infty {\rm d}\tau \phi(\tau),
\end{equation}
assuming this integral exists.
\section{Examples}
In this section we give two examples of how a laser field and an atom become entangled during quantum-information processing operations. We first consider on-resonance excitation of a two-level atom with a single laser field.
\subsection{Quadrupole transition}\label{quadru}
\begin{figure}\leavevmode
\epsfxsize=8cm \epsfbox{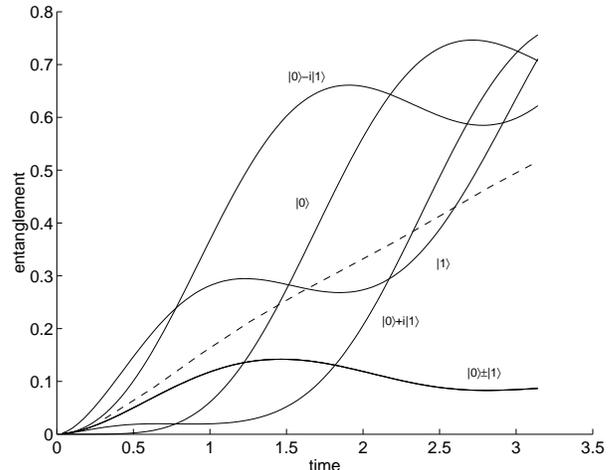} \caption{Entanglement as a function of scaled time $\tau=g|\alpha|\tilde{t}$. 
For the five solid curves the corresponding initial atomic state is indicated in the figure, and the dashed curve gives the entanglement averaged over all possible initial atomic states.}
\label{ent1} \end{figure}

In the case of on-resonance excitation with a coherent state of large amplitude there is a simple relation between the total interaction time $T$ and the type of operation we wish to perform. For instance, a NOT operation corresponds to a time $T$ such that $g|\alpha|T=\pi/2$. This leads us to define a scaled time variable $\tau$ according to
\begin{equation}
\tau=g\tilde{t}|\alpha|,
\end{equation}
such that a NOT operation  corresponds to $\tau=\pi/2$.
We first give an example of the dependence of the atom-field entanglement on the initial atomic state. In Fig.~\ref{ent1} we use a coherent state with a real amplitude $\alpha=\sqrt{10}$ and plot $E(\theta,\varphi)$ for various initial states. Note that the absolute value of the phase $\varphi$ has no physical meaning and only its relative value compared to the phase arg$(\alpha)$ of the coherent state matters. The Figuire shows that for an atom initially in the ground state the atom-field entanglement is relatively small for short times.  This is a direct consequence of the coherent state being an eigenstate of the annihilation operator $a$. Namely, for an atom in the ground state, the term proportional to $\sigma^{\dagger} a$ is the only term in the Hamiltonian giving rise to nontrivial evolution, and this term alone does not entangle atom and laser field. 
Conversely, the atom-field entanglement rises the most quickly for an atom starting in the excited state.

\begin{figure}\leavevmode
\epsfxsize=8cm \epsfbox{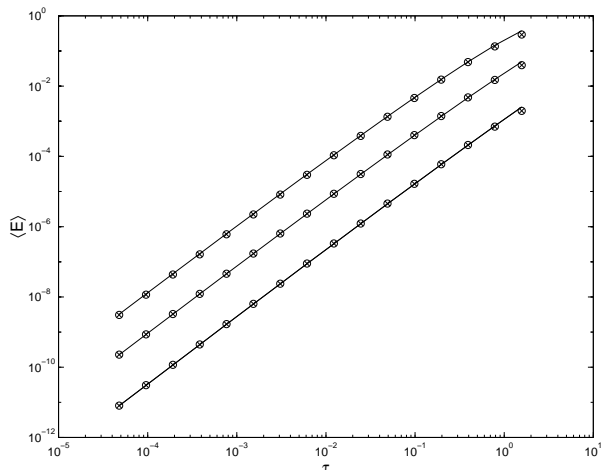} \caption{Average entanglement as a function of scaled time $\tau$ for different values of the average photon number $\bar{n}=|\alpha|^2$; from top to bottom curves we have $|\alpha|^2=2^3,2^7,2^{12}$. The filled circles indicate numerical results for times $\tau_n=2^{-n}\pi$ for $n=1\ldots 16$, while the solid curves correspond to the approximate analytic solution (\protect{\ref{entQ}}).}
\label{ent1all} \end{figure}

In Fig.~\ref{ent1all} we plot the average entanglement $\langle E\rangle$ as a function of time $\tau$ for a resonant interaction. 
As expected the entanglement decreases with photon number and increases with $\tau$ (up to the point where atom and field become almost completely entangled). 
We can in fact get an analytic result for $\tau\ll 1$. 
We expand the evolution operator to first order in $g\tilde{t}|\alpha|=\tau$ by
\begin{equation}
1-ig\tilde{t}(a^{\dagger}\sigma^-+\sigma^{\dagger}a) 
\end{equation}
for $\Delta=0$, which gives rise to an approximate evolution of the form
\begin{eqnarray}
&&(|\beta_0|0\rangle+\beta_1|1\rangle)\otimes |\alpha\rangle\mapsto\nonumber\\
&&(|\beta_0|0\rangle+\beta_1|1\rangle)\otimes |\alpha\rangle
-igt\alpha((|\beta_0|1\rangle+\beta_1|0\rangle)\otimes |\alpha\rangle)\nonumber\\
&&-it\beta_1|0\rangle\otimes(a^{\dagger}|\alpha\rangle-\alpha|\alpha\rangle),
\end{eqnarray}
where for simplicity we take $\alpha$ real. The field state appearing in the last term is orthogonal to $|\alpha\rangle$ and is normalized, so that it becomes straightforward to trace out the laser field and obtain the reduced atomic density matrix. 
Its two eigenvalues $\lambda_{\pm}$ are 
(substituting $|\beta_1|=|\sin(\theta/2)|$)
\begin{equation}
\lambda_\pm\approx\frac{1}{2}\pm \frac{1}{2}\sqrt{1-4\tau^2\sin^4(\theta/2)/|\alpha|^2}.
\end{equation}
With these eigenvalues we find the entanglement (\ref{entg}) as a function of $\theta$, which in turn yields the average entanglement (\ref{enta}) by averaging over $\theta$,
\begin{equation}\label{entQ}
\langle E\rangle(\tau)\approx \left(\frac{1}{3}+\frac{2}{9\ln 2}\right)\frac{\tau^2}{|\alpha|^2}
-\frac{1}{3}\frac{\tau^2}{|\alpha|^2}\log_2\left(\frac{\tau^2}{|\alpha|^2}\right).
\end{equation}
This expression describes the entanglement well even when $\tau$ is not small but $\tau^2/|\alpha|^2$ is.
\subsubsection{Entanglement in typical experiments}
So how much do atom and field become entangled in a typical quantum information processing experiment using a quadrupole transition in an ion (see, e.g., \cite{HCN})?
The answer depends on several variables. First, let's assume we are interested in performing a NOT operation, so that we fix $T$ such that $g|\alpha| T=\pi/2$.
The answer then may depend on the values of the focusing parameter $A$ of the laser field, the coupling constant $d=2\pi Q/\lambda$, the frequency $\omega_L$, and the power $P$ of the laser, which determines the flux $F$ by $F=P/(\hbar \omega_L)$.
In terms of the total interaction time $T$ we can approximate $\alpha$ by
\begin{equation}
\alpha\approx \sqrt{F/{\cal T}}T\approx \sqrt{FT},
\end{equation}
where we used that ${\cal T}$ and $T$ are of the same order of magnitude. (For instance, for a Gaussian laser pulse they would differ by a factor of $\sqrt{2}$.)
Substituting these relations into the expression for the interaction time $T$ gives
\begin{equation}
T=\frac{\pi\hbar}{d}\sqrt{\frac{\epsilon_0 cA}{2P}}.
\end{equation}
As expected $T$ increases with $A$ (weaker focusing) but decreases with power $P$ and the coupling constant $d$ (stronger coupling).
The number of photons $|\alpha|^2$ necessary to perform a $\pi$ pulse in time $T$ is
\begin{equation}
|\alpha|^2\approx FT=\frac{\pi}{\omega_L d}\sqrt{\frac{\epsilon_0 c AP}{2}}.
\end{equation}
The average entanglement (\ref{entQ}) depends only on the ratio $\tau^2/|\alpha|^2$, where here $\tau=\pi/2$.
Assuming a focusing area of $A\approx 100 (\mu{\rm m})^2$, a typical quadrupole moment of $Q\approx ea_0^2$ \cite{quadru} and a power of $P=100 \mu$W, yields for a wavelength of 730nm (corresponding to the S$_{1/2}$ to D$_{5/2}$ transition in $^{40}$Ca$^+$ \cite{HCN})
$T\approx 3.1 \mu$s and $|\alpha|^2\approx 1.1\times 10^{9}$, so that
\[
E\approx 2.2\times 10^{-8}.
\]
Decreasing the focal area $A$ will increase the amount of entanglement. But even very strong focusing to areas of size $A\approx \lambda^2$ still does not lead to large entanglement. In fact, if we make $A$ smaller by a factor 100 so that $A\approx 1 (\mu {\rm m})^2$ (although we should note that the 1-dimensional model of Eq.~(\ref{1D}) would cease to be valid for such small values of $A$) and decrease the power by a factor of 100 as well such that $T$ remains constant, the entanglement increases by about a factor of 100 to a value $E\approx 10^{-6}$ that is still very small. 
    
It is interesting to compare the smallness of the entanglement to the probability of spontaneous emission. Here the lifetime $\tau_0$ of the metastable D$_{5/2}$  state is about 1 sec. Since the interaction time is $T=3.1\mu$ sec, the spontaneous emission probability during a NOT operation is thus $p_{{\rm spon}}=T/(2\tau_0)=1.6\times 10^{-6}$ (the factor 1/2 arises since the atom spends half of the time in the excited state).

It is perhaps also interesting to compare these numbers to those for a dipole transition under similar circumstances. For instance, for the 6S$_{1/2}$ to 6P$_{3/2}$ dipole transition in Cs (at a 
wavelength $\lambda\approx 850$nm, and an upper state lifetime of $\tau_0\approx 31$ns), at the same laser power $P$ and the same focusing area $A$, one would have a duration of $T=0.46$ns for a NOT operation, an entanglement of $E=7.6\times 10^{-5}$ and a spontaneous emission probability during the NOT operation of $p_{{\rm spon}}=0.0073$. Thus for a dipole transition the decoherence effects of both spontaneous emission and entanglement are larger by more than three orders of magnitude than for a quadrupole transition.  
\subsection{Raman transition}\label{raman}
\begin{figure}\leavevmode
\epsfxsize=8cm \epsfbox{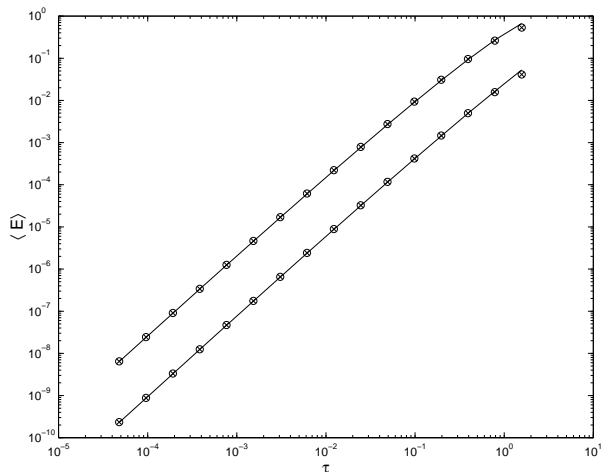} \caption{Average entanglement as a function of scaled time $\tau$ for different values of the average photon number $\bar{n}=|\alpha|^2$; from top to bottom curves we have $|\alpha|^2=2^3,2^8$. The filled circles indicate numerical results for times $\tau_n=2^{-n}\pi$ for $n=1\ldots 16$, while the solid curves correspond to the approximate analytic solution \protect{(\ref{entR})}.
}\label{meall} 
\end{figure}

We now consider a typical situation where two ground states of an atom are used as quantum bits. The two ground states are connected through a two-photon Raman transition (now dipole-allowed transitions) via an intermediate far off-resonant excited state. The detuning $\delta$ from the excited state is usually chosen much larger than the decay rate of that excited state, so that dissipation can be neglected. 
As in the previous subsection, when the bandwidths of the two fields are sufficiently small (conditions are given below) we may introduce two single-mode annihilation operators, $a_1$ and $a_2$, one for each laser field. We assume two-photon resonance so that
the effective interaction Hamiltonian takes the form
\begin{equation}
\tilde{H}(t)=\hbar \Omega\phi(t)(\sigma^{\dagger} a_1^+a_2 +\sigma a_2^+a_1 ).
\end{equation}
with
\begin{equation}
\Omega=\frac{g_1g_2}{\delta},
\end{equation}
where $g_1$ and $g_2$ are the coupling constants for the two dipole transitions, defined as in (\ref{ga}).
Similarly, $\phi(t)$ now refers to the product of two envelope functions, one for each laser field. 
For simplicity we assume $g_1=g_2\equiv g$ and the initial field states to be coherent states with equal amplitudes $\alpha$.
We introduce a scaled time variable
$\tau=\Omega|\alpha|^2 \tilde{t}$, such that a NOT operation corresponds to $\tau=\pi/2$.
For small times and large amplitudes we can approximate (as before) the two eigenvalues of the reduced atomic density matrix by
\begin{equation}
\lambda_\pm\approx\frac{1}{2}\pm \frac{1}{2}\sqrt{1-4\tau^2[\sin^4(\theta/2)
+\cos^4(\theta/2)]/|\alpha|^2}.
\end{equation}
The average entanglement after a time $\tau$ is then
\begin{equation}\label{entR}
E(\tau)\approx \left(\frac{2}{3}+X\right)\frac{\tau^2}{|\alpha|^2}
-\frac{2}{3}\frac{\tau^2}{|\alpha|^2}\log_2\left(\frac{\tau^2}{|\alpha|^2}\right),
\end{equation}
with
\[
X=-2\int_0^1{\rm d}x x (x^4+(1-x^2)^2)\log_2(x^4+(1-x^2)^2)\approx 0.3667.
\]
We plot the average entanglement as a function of the scaled time variable $\tau$
in Fig.~\ref{meall}. 
\subsubsection{Entanglement in typical experiments}
For the Raman transition too, we evaluate here to what extent atom and field become entangled in a typical experiment using Raman transitions between different ground states (see, e.g., \cite{wineland}).
We assume as before we are interested in performing a NOT operation, and we fix $T$ such that $\Omega |\alpha|^2 T=\pi/2$.
The interaction time $T$ is now given by
\begin{equation}
T=\frac{\hbar^2\delta\pi\epsilon_0 c A}{d^2P}.
\end{equation}
The interaction time increases with $A$ (weaker focusing) and $\delta$ (weaker two-photon coupling) but decreases with power $P$ and the dipole moment $d$ (stronger coupling).
The number of photons $|\alpha|^2$ in each laser beam necessary to perform a $\pi$ pulse in time $T$ is now
\begin{equation}
|\alpha|^2\approx FT=\frac{\hbar\delta\pi\epsilon_0 c A}{d^2\omega_0}.
\end{equation}
Again, this number increases with $\delta$ and $A$ for obvious reasons.
The entanglement (\ref{entR}) depends only on the ratio $\tau^2/|\alpha|^2$, where here $\tau=\pi/2$.
Assuming a focusing area of $A\approx 100 (\mu{\rm m})^2$, a typical value of $d\approx 0.2ea_0$ [using a value $ea_0$ for the dipole moment and a Lamb-Dicke factor of 0.2, see \cite{wineland}],
a power of $P=0.5$mW, and a typical value of $\delta/(2\pi)=10$GHz yields for a wavelength of 800nm $T\approx 0.41\mu$s and $|\alpha|^2\approx 8.2\times 10^8$. Just as for a quadrupole transition, such a large photon number corresponds to a small amount of entanglement
\[
E\approx 6.0\times 10^{-8}.
\]
Decreasing the focal area by a factor of 100 so that the focusing limit of $A\approx \lambda^2$ is reached, and decreasing the power $P$ by a factor 100 as well so as to keep $T$ constant, increases the entanglement by about a factor of 100, which still leaves it a small number, $E<10^{-5}$.

Finally, let us consider the effects of spontaneous emission. Due to the large detuning from the excited state the effective spontaneous emission rate $\gamma$ is small and it is given by
\begin{equation}
\gamma=\frac{\Gamma}{\delta}\Omega |\alpha|^2,
\end{equation}
with $\gamma$ the spontaneous decay rate of the excited state. The probability of spontaneous emission during a NOT operation is thus $p_{{\rm spon}}=\frac{1}{2}\gamma T=\frac{\pi}{4}\frac{\Gamma}{\delta}\approx 5\times 10^{-4}$.
Clearly spontaneous emission is a larger decoherence effect than atom-field entanglement under typical conditions.
\section{Conclusions}
We calculated the entanglement between a laser field and an atom in two different cases of interest for quantum information processing.  For two-photon ($m=2$) Raman transitions between two ground states and for single-photon ($m=1$) transitions between a ground state and a (metastable) excited state we found that the entanglement produced during a NOT operation is
\[\langle E\rangle\approx  \frac{m\pi^2}{12} \bar{n}\log_2\left(\frac{4\bar{n}}{\pi^2}\right)+{\cal O}(\bar{n})\,\,\,{\rm for} \,\bar{n}\gg 1
\]
with $\bar{n}$ the average number of photons in the laser field(s).
For typical values of laser power, focusing area etc., the entanglement between atom and field remains very small, $\langle E\rangle\approx 10^{-7},10^{-8}$.
Even for very strong focusing down to a wavelength the entanglement remains small, $\langle E\rangle\approx 10^{-5},10^{-6}$. This in some sense agrees with the conclusion arrived at in \cite{focus}, that in free space the atom-field interaction remains relatively weak so that light scattered from an atom does not contain much information about the state of the atom. This degree of entanglement is sufficiently small that errors due to this effect are below the error threshold required for fault-tolerant quantum computation \cite{preskill}.
On the other hand, it means that in practice other effects will be much more important causes of decoherence.  For instance, even in typical cases where spontaneous emission is suppressed, either by choosing a metastable state as qubit state or by detuning far from resonance with unstable excited states, the decoherence due to atom-field entanglement is smaller than that due to spontaneous emission by a few orders of magnitude.
\section*{Acknowledgements}The work of HJK was supported by the National Science Foundation, by the
Caltech MURI on Quantum Networks administered by the US Army Research
Office, and by the Office of Naval Research. 


\begin{thebibliography}{99}
\bibitem{mixed}We are aware \cite{enkfuchs} of the issue raised by K. M\o lmer in \cite{molmer} that the quantum state of a laser is in fact not a coherent state, but rather a mixture of coherent states with random phases. Here, since we average over all phases of the atomic initial state, this fact is not important. Another way to see that this point is harmless in our present discussion is the following. If we assume the initial state of laser plus medium to be of the form
\[
\sum_n \alpha_n |n\rangle \otimes |N-n\rangle,
\]
with $N$ giving the total number of initially excited atoms in the medium and $n$ the number of photons produced, then as far as the entanglement of the atom is concerned we will find the same answer if we assumed the laser field to be in a pure state 
\[
\sum_n \alpha_n |n\rangle.
\]
\bibitem{quadru}To have an independent check for this value of the quadrupole moment, begin with an expression for the radiative lifetime of a
dipole-allowed transition, namely 
\[
\tau _{0}=\frac{3\pi \epsilon _{0}\hbar c^{3}}{d^{2}\omega _{0}^{3}}
\]
and replace $d\rightarrow kQ$. The lifetime for
the $D_{5/2}\rightarrow S_{1/2}$ transition in $^{40}$Ca$^{+}$
is $\tau _{0}\approx 1$sec. Plugging this value into the above expression
for $\tau _{0}$ (with transition wavelength $\lambda _{0}=730$nm) and
solving for $Q$ gives $Q=3.9\times 10^{-40}$Coul-m, which agrees quite
nicely with the value $Q_{0}=ea_{0}^{2}\approx 4.5\times 10^{-40}$Coul-m.
\end{thebibliography}
\end{document}